\documentclass[reprint,aip,numerical]{revtex4-1}
\usepackage{graphicx}%

\begin{document}


\title{UV-photoelectric effect for augmented contrast and resolution in electron microscopy} 



\author{Gediminas Seniutinas}
\email[]{gdseniutinas@gmail.com}
\affiliation{Faculty of Science, Engineering and Technology,
Swinburne University of Technology, John st., Hawthorn, Vic. 3122,
Australia}

\author{Armandas Bal\v{c}ytis}
\affiliation{Faculty of Science, Engineering and Technology,
Swinburne University of Technology, John st., Hawthorn, Vic. 3122,
Australia} \affiliation{Institute of Physics, Center for Physical
Sciences and Technology, 231 Savanori\c{u} Avenue, LT-02300
Vilnius, Lithuania}

\author{Saulius Juodkazis}
\email[]{sjuodkazis@swin.edu.au}
\affiliation{Faculty of Science, Engineering and Technology,
Swinburne University of Technology, John st., Hawthorn, Vic. 3122,
Australia}


\date{\today}

\begin{abstract}
A new tool providing material contrast control in scanning
electron microscopy (SEM) is demonstrated. The approach is based
on deep-UV illumination during SEM imaging and delivers a novel
material based contrast as well as higher resolution due to the
photoelectric effect. Electrons liberated from illuminated sample
surface contribute to the imaging which can be carried out at a
faster acquisition rate, provide material selective contrast,
reduce distortions caused by surface charging, and can substitute
metal coating in SEM. These features provide high fidelity SEM
imaging and are expected to significantly improve the performance
of electron beam instruments as well as to open new opportunities
for imaging and characterization of materials at the nanoscale.
\end{abstract}


\maketitle 

\section{Introduction}

Breakthrough in nanotechnology over the last decades and its
current progress are highly related to advances in fabrication and
characterization techniques based on electron
beams~\cite{li2004engineering, muller2009structure, yu2011light,
manfrinato2013resolution}. Scanning and transmission electron
microscopies (SEM \& TEM) provide imaging capabilities down to
atomic resolution and enable to identify materials and determine
their physical properties at the nanoscale~\cite{krivanek2010atom,
wang2000transmission, gericke2008high}. Signals from the electron
beam generated secondary and scattered primary electrons are used
to build up electron micrographs, thus spatial imaging resolution
mostly depends on the probe spot size and interaction volume of
electrons in the sample\cite{joy1984beam}. As the spot size
defines the resolution, the beam current determines the time
needed to acquire an image. An ideal electron source would provide
a high brightness single dot beam, however, in reality a trade-off
between the spot size and the current of the beam has to be made.

Schottky electron emitters, cold field-emission guns and emerging
carbon nanotube electron sources are used in state-of-the-art
electron microscopes and provide high brightness with relatively
small energy spread of the electron beam~\cite{orloff1989survey,
van2001reduced, de2002high}. Nevertheless, it still takes a few
minutes to take a good quality image using the lowest currents and
the sample might move or change during this period resulting in a
blurred image. At higher currents resolution is diminished not
only because of a larger spot size but, also, due to surface
charging in insulating or low electrical conductivity samples.
Polymer coatings (e-spacer)~\cite{sun2012high} as well as highly
conductive Pd/Pt sputtered films 1-3 nm in thickness are solutions
currently widely used to mitigate surface charging. However, as
demand for higher resolution approaching 1-2 nm is commonplace in
materials science and electronics~\cite{gupta14,IBMpr}, coatings
of such thickness are already not acceptable. Moreover, samples
can hardly be used in further processing steps after imaging as
the coatings cannot be removed easily. Alternative approach to
minimize surface charging effects is to use low vacuum imaging
mode, but in this case imaging resolution is
worsened~\cite{donald2003use, thiel2005secondary}.

Here, we exploit the photoelectric effect to demonstrate a novel
approach to address the aforementioned challenges and further
improve the capabilities of electron microscopy. It is shown that
light emitting diodes (LEDs) at the deep-UV (DUV) spectral range
provide capability to generate electrons from the sample?s surface
leading to a faster image acquisition, reduced surface charging
effects and improved contrast which, in addition, becomes material
dependent.

\section{Experimental}

\begin{figure}[tb]
 \centering
\includegraphics[width=\columnwidth]{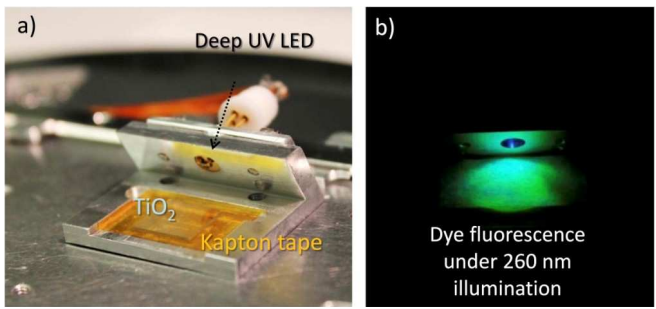}
\caption{a) A homemade sample sub-stage for mounting light
emitting diodes. Insulating Kapton tape is used in between the
sub-stage and a sample to increase the surface charging effect; b)
Powder fluorescence under illumination of 260 nm wavelength light
showing the illumination intensity distribution.} \label{f-led}
\end{figure}

Influence of the photoelectric effect on SEM imaging quality was
studied using DUV LEDs emitting at 250-280 nm wavelengths (SETi,
Ltd.). The LEDs were placed in a homemade sample sub-stage
(Fig.~\ref{f-led}(a)). The sub-stage was mounted on a sample
holder of an electron beam lithography tool (Raith
$150^{\text{TWO}}$) which can also be used as a high resolution
SEM instrument. To minimize reflection (increase absorption) the
angle of DUV light incidence on the sample was chosen $\theta \sim
60^{\circ}$ which is close to the Brewster angle. Power supply for
the LEDs was connected via vacuum chamber feed-through contacts
and controlled from the outside.

The LEDs used did not have any collimation/focusing optics on the
emitting chip, hence, the sample illumination was not uniform over
its area and it was difficult to calculate illuminance at
particular imaging spots. The illumination profile is shown in
Fig.~\ref{f-led}(b) where paper with a fluorescent powder (UVSWG,
absorption peak at 260 nm, emission peak at 525 nm) is placed
instead of a sample and illuminated under 260 nm light. SEM
imaging was carried out in the highest illuminance area. The UV
exposure was proportional to the LED current which was used as a
measure of the dose. All the experiments were carried out at
maximum DUV illumination power of several $\mu$W at a current of
20 mA. For the SEM imaging experiments titania (TiO$_{2}$) and
silicon were used as substrates for test sample fabrication.

Structures of 50 nm thick metals with different electron work
functions, $\phi$, (Au (5.1-5.47 eV), Ti (4.33 eV) and Al
(4.06-4.26 eV)) were patterned onto the substrates using a
standard electron beam lithography and lift-off process. Titania
is a semi-insulating material with the electron work function of
4.9 eV~\cite{borodin2011characterizing}, whereas Si is conductive
with the work function of 4.60-4.85 eV. Silicon is a suitable
substrate for SEM imaging but needs to be well grounded to
facilitate removal of scattered and secondary electrons that are
dispersed over the surface. For dielectric and semiconducting
surfaces with low electrical conductivity the Maxwell relaxation
time is long and strong charging occurs causing image distortions
even though the sample is well grounded; the relaxation time is
defined by the electric susceptibility $\varepsilon$ and
conductivity $\sigma$ as $\tau = \sigma / \varepsilon$. In our
experiments, an insulating Kapton tape was placed between the
samples and the sub-stage (Fig.~\ref{f-led}(a)) to prevent removal
of electrons through grounding, resulting in an increase of
surface charging, hence to better reveal a charge removal effect
by DUV co-illumination. The electron beam current was set to 0.4
nA and accelerating voltage of 10 kV was used for all imaging
experiments, which also were in favour of an increase in surface
charging.

\begin{figure*}[t]
 \centering
\includegraphics[width=\textwidth]{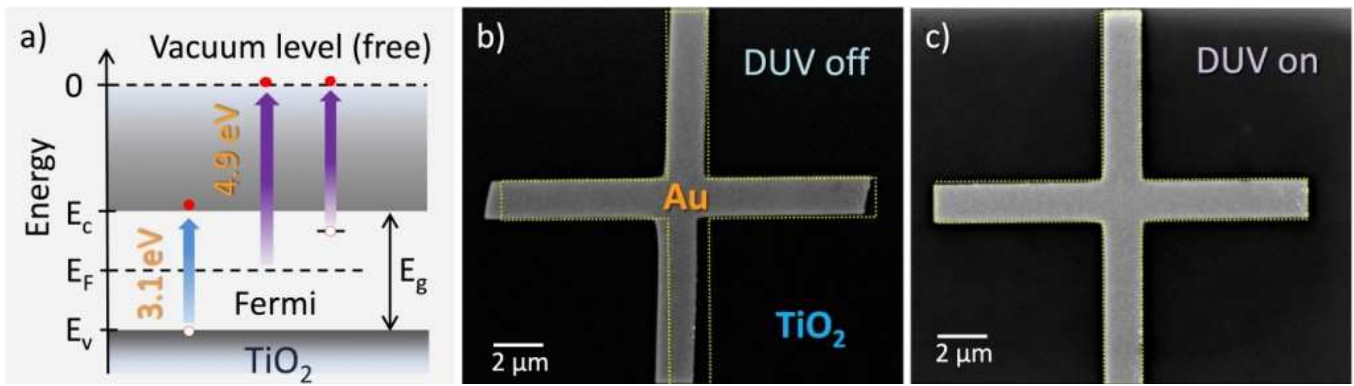}
\caption{Surface charging control: a) Schematic which shows the
energy levels of the titania substrate used; b) and c) show SEM
images of Au cross taken without and with DUV illumination
respectively. The illumination wavelength was $\sim$260 nm.}
\label{f-sem}
\end{figure*}

\section{Results and discussion}

Surface charging is the main cause of SEM image distortions when
insulating materials are being imaged. It becomes a challenge to
image nanometer sized structures or estimate a thickness of thin
deposited layers in cross sectional imaging when conductive
coatings cannot be applied.

First, surface charge removal has been tested for TiO$_{2}$
substrate. The 260 nm wavelength LED light has slightly lower
photon energy than the work function of titania: 4.9 eV or
equivalent to a 250 nm wavelength. However, electrons from any
surface defects are efficiently excited at 260 nm as it was
demonstrated in the case of Ga$^{+}$-ion milling under DUV
illumination~\cite{13lpr1049}. SEM imaging has been tested under
DUV co-illumination which is strongly absorbed at the surface.
Surface defects in semiconducting and dielectric materials can be
represented as defect levels in the band diagram, e.g., for the
electron donor/trap typical depths are 0.1-0.3 eV below the
conduction band (Fig.~\ref{f-sem}(a)). Liberation of electrons to
the free vacuum level then requires correspondingly smaller photon
quantum energies (slightly longer wavelengths) as compared with
the work function, $\phi$.

\begin{figure}[b]
 \centering
\includegraphics[width=\columnwidth]{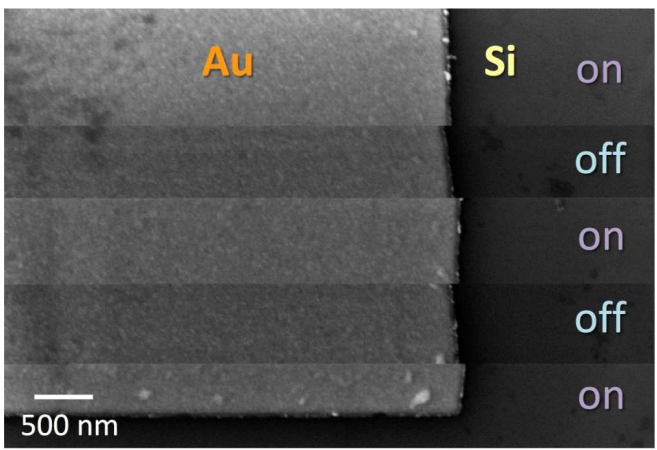}
\caption{SEM image taken by alternatingly switching DUV
illumination on and off during image acquisition.} \label{f-rt}
\end{figure}

The prepared samples were placed on the sub-stage as shown in
Fig.~\ref{f-led}(a) and imaged in the SEM. A secondary electron
detector operating at line averaging mode was used to better
visualize image shifts resulting due to surface charging. Total
scan time of one line was 26.3~ms and it took 20.2~s to acquire an
entire 1024$\times$768 pixel image. The selected mode minimizes
distortions in a single line as the line is scanned rapidly, but
overall image obtained upon combining all the individual lines is
distorted due to the charge accumulation induced drift. Images
taken without and with DUV illumination are shown in
Fig.~\ref{f-sem}(b) and (c), respectively. Without the DUV
exposure the image drifts and becomes distorted; a contour of the
actual patterned cross is indicated by the yellow dotted outline.
The drift is avoided under DUV resulting in unequivocal imaging.
The photoelectric effect causes any accumulated electrons to
escape from the sample surface, hence, the drift is eliminated.
Wide area illumination favors removal of charge gradients, hence,
reduces currents, and it does not perturb the imaging electron
beam. Figure~\ref{f-rt} shows SEM image of a 50 nm thick gold
structure on a silicon substrate. Deep UV illumination at 260 nm
wavelength was alternatingly switched on and off during the image
acquisition, which allowed to monitor changes induced by DUV
exposure \textit{in situ}. Clearly, the image becomes brighter
when the illumination is on due to additional electrons liberated
from the sample. Image drift can be qualitatively evaluated based
on the edge angle changes in the "on" and "off" cases. The edge is
straight under DUV exposure and starts to drift once the light is
switched off. A clear correlation between the image drift and LED
current has been confirmed.

Another advantage of DUV illumination during SEM image acquisition
is the new ability to exploit material contrast. In most cases
backscattered electron detection is used to obtain material
contrast in SEM. Material contrast imaging with a high resolution
is in demand for failure analysis in
electronics~\cite{ratchev2004reliability, coyle2009influence},
mineralogy~\cite{gottlieb2000using} and materials
nano-engineering~\cite{hu2006photosensitive, bruet2008materials,
hibi2014selective}. Using DUV illumination with photon energies
close to the work function, materials with differing values of the
work function are readily revealed. Under UV exposure the contrast
between different materials appears not only due to disparate
rates of secondary electron generation, but because of differences
in the work function as well.

This newly proposed material contrast dependence on the work
function was tested for Ti, Au and Al structures on a TiO$_{2}$
substrate. Quarter-pie patterns of each metal were fabricated on
the same substrate and spaced within a few microns to fit into one
field of view during SEM imaging. Elemental analysis was first
carried out by SEM energy dispersive x-ray spectroscopy (EDX). The
elemental map identifying all three metals is provided in
supplementary material Fig. S1. Oxygen signal has been observed
over the Al structure indicates that the surface was oxidized and
the work function has been correspondingly altered.

\begin{figure*}[tb]
 \centering
\includegraphics[width=\textwidth]{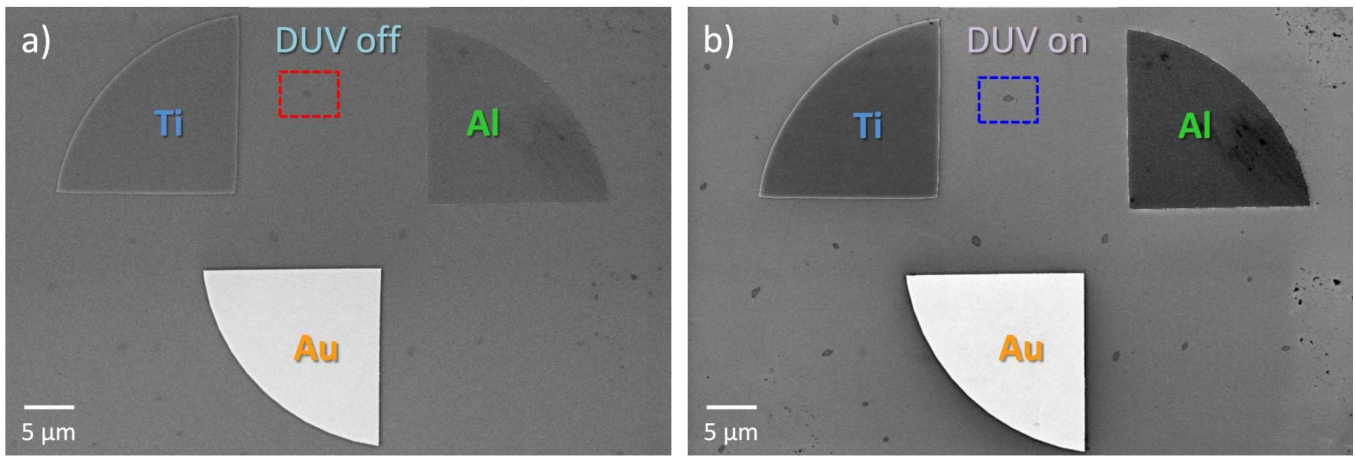}
\caption{SEM imaging of the work function under illumination at
$\sim$260 nm wavelength. Darker appearing metal has a smaller work
function. A gray scale analysis of the marked areas is presented
in the supplementary material Fig. S2.} \label{f-smile}
\end{figure*}

The quarter-pie structures were imaged with and without DUV light
(Fig.~\ref{f-smile}). The acquisition contrast was optimized for
the best imaging without DUV illumination. All SEM settings were
kept the same when the DUV was switched on. Aluminium under
illumination, due to its lowest work function
($\varphi_{Al}<\varphi_{Ti}<\varphi_{Au}$), is stripped of the
largest amount of electrons and appears the darkest. Materials
most deficient in electrons look darker because secondary
electrons require more energy to escape them as surface potential
barrier becomes higher.

A generic scaling of light - matter interaction defined by the
cross section, $\sigma_{c}$, which determines the probability of
photoelectric effect depends on the atomic number, $Z$, of the
target material and the photon energy, $E = h \nu$, as $\sigma_{c}
\sim Z^{n}/(h \nu)^{3}$ ~\cite{Siegbahn} where the exponent $n$ is
bound between $4 < n < 5$. Hence, a stronger light matter
interaction is expected for heavier elements, in our case $Z_{Au}
>Z_{Ti} >Z_{Al}$. However, the contrast dependence is more
complex.

With the increased contrast between different materials, a higher
resolution could be achieved. As shown in Fig.~\ref{f-smile}, the
residues of poly(methyl methacrylate) resist with the work
function of  $\varphi =$ 4.68~eV~\cite{van2009properties} are
resolved much better under the DUV illumination. A gray scale
analysis of areas marked in Fig.~\ref{f-smile} showed a wider
dynamic range as well as larger changes in pixel signal values
around the residue edges for the image taken under the
illumination. The analysis graph is presented in supplementary
material Fig. S2. The technique is highly appealing for
nanomaterial research as better contrast due to material response
yields an increase of the resolution when different materials are
being imaged.

Combination of several wavelength LEDs and their availability at
higher luminous fluences are expected to create new solutions for
advanced SEM imaging. Deep-UV illumination incorporated into high
resolution SEM microscopes could substitute the metal coating of
samples for SEM imaging. This ?optical coating? will simplify and
accelerate SEM imaging making it more flexible and compatible with
production line inspection without damaging a chip; of special
note is that the inspected chip can go into further processing
steps and subsequently be imaged for a better process control
feedback. Deep-UV co-illumination simultaneous with SEM imaging
demonstrated here for the first time makes this technology also
highly appealing for imaging of various nanomaterials. The
required instrumentation to realize a LED light flood gun in a
standard SEM microscope is straightforward. One could foresee DUV
LEDs mounted directly inside a microscope?s vacuum chamber rather
than used on a sample holder as in this proof-of-the-principle
study.

\section{Conclusions}

Deep-UV co-illumination during SEM imaging creates a new method
for contrast control via the work function of material being
imaged. Moreover, SEM image distortions due to surface charging
are mitigated via removal of strong charge gradients by DUV light.
This method can strongly advance material research especially
where high resolution imaging of dielectrics and metals placed at
nanoscale proximity is required as well as for probing
bio-materials where use of conductive coatings is prohibitive.
Deep-UV illumination is expected to advance nanofabrication
methods based on electron and ion beam techniques and substitute
metal coating in imaging.

\begin{acknowledgments}

SJ is grateful for partial support via the Australian Research
Council DP130101205 Discovery project.

\end{acknowledgments}



%
%

%



%

\newpage
\section{Supplement}
\begin{figure}[h]
 \centering
\includegraphics[width=\columnwidth]{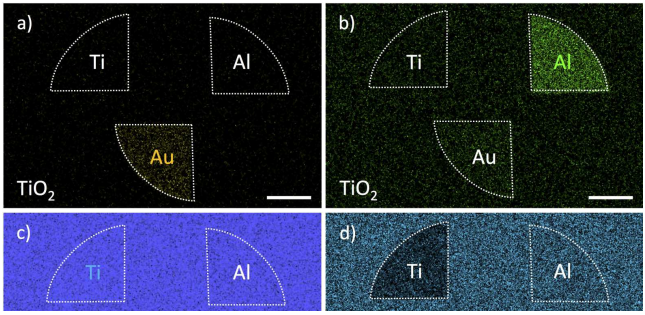}
\caption{Elemental mapping of the test sample with titanium,
aluminium and gold structures on titania substrate. EDX maps show
the following elements: a) gold, b) aluminium, c) titanium, d)
oxygen. Oxygen signal on the Al structure indicates that the
surface was oxidized and the work function has been
correspondingly altered. On the other hand, Ti over Al is observed
because low atomic mass Aluminium is only 50 nm thick, therefore
electrons from the scanning beam with energy of 30 keV easily pass
through it and excite Ti in the TiO$_2$ substrate. Scale bar is
10~$\mu$m.} \label{f-rt}
\end{figure}

\begin{figure}[h]
 \centering
\includegraphics[width=\columnwidth]{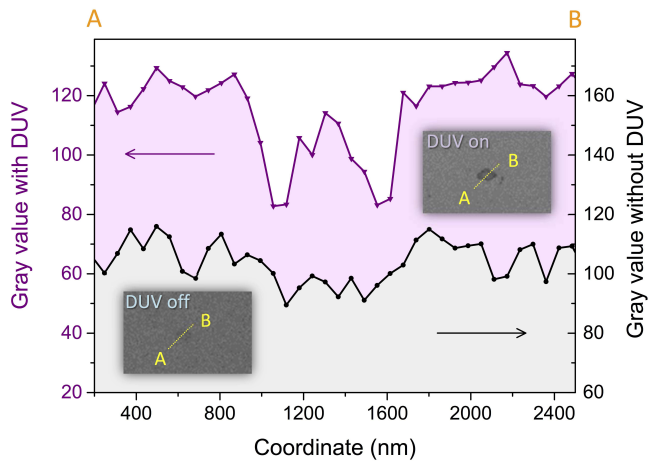}
\caption{Gray values of areas marked in Fig. 5 in the main text
are analyzed along AB line for the images taken with and without
DUV illumination. The dynamic range in the image taken under UV
illumination is broader and smaller features can be resolved as
shown in the insets.} \label{f-rt}
\end{figure}

\clearpage

\end{document}